\documentclass[conference]{IEEEtran}
\IEEEoverridecommandlockouts

\usepackage{glossaries}
\newacronym{os}{OS}{Operating System}
\newacronym{nyi}{NYI}{Not Yet Implemented}
\newacronym{dei}{DEI}{Department of Informatics Engineering}
\newacronym{AI}{AI}{Artificial Intelligence}
\newacronym{BLEU}{BLEU}{Bilingual Evaluation Understudy}
\newacronym{ROUGE}{ROUGE}{Recall-Oriented Understudy for Gisting Evaluation}
\newacronym{METEOR}{METEOR}{Metric for Evaluation of Translation with Explicit ORdering}
\newacronym{ML}{ML}{Machine Learning}
\newacronym{NLP}{NLP}{Natural Language Processing}
\newacronym{ICL}{ICL}{In-Context Learning}
\newacronym{RAG}{RAG}{Retrieval Augmented Generation}
\newacronym{LSTM}{LSTM}{Long Short-Term Memory}
\newacronym{RNNs}{RNNs}{Recurrent Neural Networks}
\newacronym{BERT}{BERT}{Bidirectional Encoder Representations from Transformers}
\newacronym{MLM}{MLM}{Masked Language Modeling}
\newacronym{NSP}{NSP}{Next Sentence Prediction}
\newacronym{ASTs}{ASTs}{Abstract Syntax Trees}
\newacronym{MBPP}{MBPP}{Mostly Basic Python Problems}
\newacronym{RLHF}{RLHF}{Reinforcement Learning from Human Feedback}
\newacronym{GenAI}{GenAI}{Generative Artificial Intelligence}
\newacronym{CoT}{CoT}{Chain-of-Thought}
\newacronym{LLOC}{LLOC}{Logical Lines of Code}
\newacronym{SLOC}{SLOC}{Source Lines of Code}
\newacronym{LOC}{LOC}{Lines of Code}
\newacronym{MoE}{MoE}{Mixture of Experts}
\newacronym{GMM}{GMM}{Gaussian Mixture Models}

\newglossaryentry{LLM}
{
  name={LLM},
  description={Large Language Model},
  first={Large Language Model (LLM)},
  plural={LLMs},
  descriptionplural={Large Language Models},
  firstplural={Large Language Models (LLMs)}
}

\newacronym{ODC}{ODC}{Orthogonal defect classification}

\newif\ifshowcomments
\showcommentstrue

\newcommand{\rn}[1]{\ifshowcomments\textcolor{black}{#1}\fi}

\usepackage{amsmath,amsfonts}
\usepackage{algorithmic}
\usepackage{listings}
\usepackage{graphicx}
\usepackage{comment}
\usepackage{subcaption}
\usepackage{cite}
\usepackage{csquotes}
\usepackage{textcomp}
\usepackage{booktabs}
\usepackage{xurl}
\usepackage{multirow}
\usepackage{xcolor}
\usepackage{makecell}
\usepackage{url}
\def\BibTeX{{\rm B\kern-.05em{\sc i\kern-.025em b}\kern-.08em
    T\kern-.1667em\lower.7ex\hbox{E}\kern-.125emX}}

\lstdefinestyle{smallcpp}{
    language=C++,
    basicstyle=\ttfamily\fontsize{6pt}{6pt}\selectfont,
    backgroundcolor=\color{gray!10},
    frame=single,
    rulecolor=\color{gray},
    breaklines=true,
    showstringspaces=false,
    tabsize=2,
    captionpos=b,
    numbers=left,
    numberstyle=\tiny\color{gray},
    numbersep=5pt,
    xleftmargin=5pt,  
    xrightmargin=5pt
}

\lstdefinestyle{smalljava}{
    language=Java,
    basicstyle=\ttfamily\fontsize{6pt}{6pt}\selectfont,
    backgroundcolor=\color{gray!10},
    frame=single,
    rulecolor=\color{gray},
    breaklines=true,
    showstringspaces=false,
    tabsize=2,
    captionpos=b,
    numbers=left,
    numberstyle=\tiny\color{gray},
    numbersep=5pt,
    xleftmargin=5pt,  
    xrightmargin=5pt
}

\usepackage{enumitem}
\setlist[itemize]{noitemsep, topsep=-2.0pt}
\usepackage{eso-pic}
\begin{document}
\AddToShipoutPictureFG*{%
  \AtPageUpperLeft{%
    \raisebox{-1.0cm}{%
      \makebox[\paperwidth][c]{%
        \parbox{0.9\paperwidth}{%
          \centering
          \small\itshape
          Preprint version of a manuscript accepted for publication in the
          \textit{37th IEEE International Symposium on Software Reliability
          Engineering (ISSRE) 2026}, Limassol, Cyprus.
        }%
      }%
    }%
  }%
}

\bstctlcite{IEEEexample:BSTcontrol}

\title{Unreliable in Practice? A Comprehensive Study of Errors in LLM-Generated Code\\
\thanks{This work is funded by national funds through FCT – Foundation for Science and Technology, I.P., within the scope of the research unit UID/00326 - Centre for Informatics and Systems of the University of Coimbra, https://doi.org/10.54499/UID/00326/2025. This work used computing resources provided through the Google Cloud Research Credits program with the
award GCP19980904 and AWS Cloud Credits for Research provided by Amazon Web Services.}
}

\author{%
  Rodrigo Pato Nogueira\IEEEauthorrefmark{1}\IEEEauthorrefmark{2},
  Marco Vieira\IEEEauthorrefmark{2},
  João R. Campos\IEEEauthorrefmark{1}\\[0.8ex]
  \IEEEauthorblockA{%
    \makebox[\textwidth][c]{%
      \begin{minipage}[t]{0.5\textwidth}\centering
        \IEEEauthorrefmark{1}\textit{University of Coimbra},
        \textit{CISUC/LASI, DEI} \\
        Coimbra, Portugal \\ jrcampos@dei.uc.pt
      \end{minipage}\hfill
      \begin{minipage}[t]{0.5\textwidth}\centering
        \IEEEauthorrefmark{2}\textit{University of North Carolina at Charlotte} \\
        Charlotte, NC, USA \\
        rpatodec@charlotte.edu, marco.vieira@charlotte.edu
      \end{minipage}%
    }%
  }%
}

\maketitle

\begin{abstract}
\glspl{LLM} are being widely used for coding, with reports indicating that AI now generates an increasing share of production code. Studies show that \glspl{LLM} can significantly improve developer productivity, yet they still struggle with more complex coding tasks. Just as understanding error modes in human-written code has been central to improving software quality, identifying and characterizing the errors in \gls{LLM}-generated code is critical for setting realistic expectations and designing mitigation strategies.
Prior research has been limited in scope, often focusing on a single language, a small number of problems, or a limited selection of models. As a result, there is still no comprehensive understanding of which errors are common and which are specific to certain models or languages.
To address these gaps and develop a deeper understanding of the quality of \gls{LLM}-generated code, we analyzed a corpus of 86,726 code samples that contained compilation or runtime errors. These samples were generated by seven \glspl{LLM} across four compiled languages. We classified errors by their underlying causes using an \gls{LLM}, manually validated these classifications, and performed a comparative analysis.
This labeled data is then used to measure error prevalence by model, language, and problem difficulty, to identify common error patterns. 
Results show that, although error types vary strongly across languages and models, even the largest models frequently make simple mistakes. We also observe that generated code often omits basic input validation or memory-safety checks, which can lead to overflows, resource exhaustion, or other reliability/security issues.
\end{abstract}


\glsresetall

\section{Introduction}

The use of \glspl{LLM} for code generation has grown rapidly, with up to 30\% of Microsoft’s code now AI-generated~\cite{microsoft2023ai} and estimates suggesting this could reach 95\% in the near future~\cite{perkel2025microsoftcto}. In addition, LLMs have been shown to significantly improve developer productivity, with studies reporting speedups of up to 55.8\%~\cite{peng2023impactaideveloperproductivity, shihab2025copilotstudents}.

Despite their widespread adoption and productivity benefits, \glspl{LLM} have significant limitations, showing weak performance on harder problems~\cite{hendrycks2021APPS, nogueira2025Beyond}. These limitations, along with their rapidly increasing use in production environments, make it essential to understand the types of errors these models produce. Such understanding helps developers set realistic expectations, design mitigation strategies, and craft more effective prompts.
However, existing studies on common errors in \gls{LLM}-generated code remain limited, often covering only one programming language (typically Python)~\cite{liu2024exploring, wang2025towards, song2023empirical, tambon2025bugs}, analyzing a small set of problems~\cite{wang2025towards, song2023empirical, tambon2025bugs} and code samples~\cite{liu2024exploring, wang2025towards, song2023empirical, tambon2025bugs, liu2024no, chen2024deep}, and examining code from only one~\cite{liu2024no, mo2025assessing} or a few models~\cite{chen2024deep}. 
These limitations can overlook errors that emerge only at scale or across different programming paradigms. Moreover, the heavy focus on Python restricts the range of errors observed, as its dynamic, interpreted nature prevents many low-level errors, such as those related to manual memory management, which are common in compiled languages. 

Just as the systematic study of errors in human-written code has been key to advancing software engineering practices, a comparable analysis of error modes across \gls{LLM}-generated code and programming languages is essential to understand their limitations and guide the development of more reliable and secure AI-assisted programming systems.

This work addresses these gaps by using a large, multi-model corpus of 86,726 code samples that resulted in compilation or runtime errors. We focus on these error types because they come with explicit diagnostic information (e.g., compiler messages or runtime exceptions), making them easier to identify and analyze consistently at scale. In contrast, incorrect outputs or inefficient solutions require reasoning about the intended behavior of the program, which is more subjective and harder to assess reliably across many samples.

The samples were generated by seven LLMs using both baseline and \gls{CoT} prompting across four compiled languages. We focus on compiled languages as they provide clearer distinctions between error types (compilation/runtime), along with explicit diagnostic messages; in interpreted languages, these distinctions are less clear, making consistent comparison more difficult. The dataset also includes additional samples generated after models received error feedback. Using an LLM-based classifier, all 86,726 compilation and runtime errors were categorized into subtypes based on two taxonomies for errors in \gls{LLM}-generated code adapted from prior work~\cite{liu2024no}. We then (1) measure overall and per-model/language error prevalence to identify trends, (2) evaluate which errors are corrected after feedback and which persist, and (3) analyze the impact of \gls{CoT} prompting on the error distribution. Unlike prior studies, our analysis considers multiple languages, models, problem difficulties, and prompting strategies, as well as iterative feedback, providing a more comprehensive characterization of \gls{LLM} code-generation errors. To the best of our knowledge, this is the first large-scale, cross-language study spanning multiple languages, models, prompts, and iterative feedback.

Results show that the kinds of errors made by \glspl{LLM} differ across programming languages and models, reflecting differences in the amount and quality of training data available for each language, as well as the challenges specific to each language. Yet, despite these differences, all models still produce a substantial number of fundamental mistakes, such as undeclared variables, missing imports, or simple type mismatches. In addition, we find that the generated code often lacks basic input checks and memory-safety safeguards (e.g., unbounded reads or unsafe memory operations), which can ultimately compromise system dependability and introduce vulnerabilities. Notably, these issues persist even in bigger commercial models such as GPT-4.1-mini, indicating that such fundamental reliability problems are not confined to smaller open-source models. These types of issues are particularly concerning, as they are more difficult to identify and have the potential for causing system failures or security vulnerabilities, and therefore have a significantly higher impact than purely functional mistakes. This highlights critical limitations that may lead to unreliable and insecure systems. In this study, we evaluate LLMs in a competitive-programming setting without IDE assistance, compiler or runtime feedback beyond standard diagnostics, or human-in-the-loop intervention; under these conditions, the results indicate that current models are still far from being suitable for reliable standalone code generation, especially in dependability- or security-sensitive settings.

To support reproducibility, we make publicly available the full experimental pipeline, including the code used to prompt the models, perform error classification, and compute all reported metrics, together with the resulting per-sample error labels, at: \url{https://tinyurl.com/UnreliableInPractice}.

In summary, the main contributions of this paper are:

\begin{itemize}[leftmargin=*]

\item The largest analysis of errors in \gls{LLM}-generated code to date, covering 86,726 compilation and runtime errors across seven models and four compiled programming languages.
\item A detailed characterization of error types, showing how they differ across models, languages, and problem difficulty.
\item The first evaluation of the impact of iterative feedback across multiple models and languages, identifying which error categories are more or less likely to be corrected.

\end{itemize}

The remainder of this paper is organized as follows. Section~\ref{sec:background} presents the background and related work, Section~\ref{sec:data} describes the experimental setup, Section~\ref{sec:results} presents and discusses the results, and Section~\ref{sec:ttv} addresses the threats to validity. Finally, Section~\ref{sec:conclusion} concludes the study.

\section{Background and Related Work}\label{sec:background}

Recently, there have been significant advances in \glspl{LLM}, and these models are now widely used in everyday developer workflows for a range of tasks \cite{openai2024gpt4technicalreport, liu2024deepseek}. One of the clearest examples is text-to-code generation: given a natural-language description, modern models can produce runnable implementations. \glspl{LLM} trained on large code corpora learn idioms, common APIs, and repository conventions, enabling them to generate functional code with little or no manual editing \cite{zhu2024deepseek, hui2024qwen25codertechnicalreport, fried2023incodergenerativemodelcode, chen2021evaluatinglargelanguagemodels}. These capabilities are already reshaping how developers prototype, complete, and test code in practice \cite{perkel2025microsoftcto}. However, \glspl{LLM} still frequently make mistakes, ranging from obvious issues, such as calling functions before declaring them, to subtler but critical problems, including missing edge cases or generating code vulnerable to issues like integer overflows or unsafe memory use \cite{liu2024no, nogueira2025Beyond}. These latter errors are particularly concerning, as they are harder to detect and may not affect functional behavior under normal conditions, while still compromising reliability and security. Understanding these errors is essential for safely and effectively relying on \glspl{LLM}.

Program failures can happen at distinct places: code can fail to compile, crash or raise errors at runtime, produce incorrect functional outputs (failing unit tests), or violate non-functional constraints such as excessive runtime or memory use. Which of these errors appears most often depends on the implementation language. Low-level languages like C/C++ expose memory-safety and undefined-behavior faults (e.g., buffer overflows, use-after-free), whereas managed or dynamic languages such as Java and Python largely prevent or abstract away these issues through runtime checks and automatic memory management~\cite{ray2014large}. As a result, errors in these languages more often manifest as API misuse, logic errors, or missing corner cases rather than low-level memory corruption. Understanding exactly where and why code fails has long been a focus of traditional software engineering, with taxonomies such as \gls{ODC}~\cite{chillarege1992orthogonal} having been developed to categorize faults and guide diagnostics.

Several recent efforts have measured how well \glspl{LLM} perform at code generation, but they differ in what they measure and how they report results. Many benchmark-style studies focus on functional correctness, evaluating models based only on whether the generated code passes a set of unit tests or not~\cite{hendrycks2021APPS, austin2021programsynthesislargelanguage, chen2021evaluatinglargelanguagemodels}. By contrast, CodeXGLUE~\cite{lu2021codexgluemachinelearningbenchmark} emphasizes reference-based similarity metrics to quantify how close a generated solution is to a canonical implementation \cite{lu2021codexgluemachinelearningbenchmark}. Although these studies provide a good idea of how \glspl{LLM} perform in code generation, they do not give users insight into what went wrong or why the generated code was incorrect. 

\begin{table}[t]
\centering
\caption{Related Work}
\vspace{-2pt}
\label{tab:related_work}
\begin{tabular}{c|ccccc}
Paper & Langs. & Problems & Samples & Models & Prompts \\ \hline
\cite{liu2024exploring}    & 1 & 1164 & 3084 & 3 & 4  \\
\cite{wang2025towards}     & 1 & 164  & 557  & 6 & 1 \\
\cite{song2023empirical}   & 1 & 164  & ?    & 3 & 1 \\
\cite{tambon2025bugs}      & 1 & 230  & 333  & 3 & 2 \\
\cite{chen2024deep}        & 2 & 2268 & ?    & 2 & 1 \\
\cite{mo2025assessing}     & 4 & 2033 & 4191 & 1 & 1 \\
\cite{liu2024no}           & 5 & 728  & 728  & 1 & 1 \\
\textbf{Our study}         & \textbf{4} & \textbf{1651} & \textbf{86,726} & \textbf{7} & \textbf{2}
\end{tabular}
\vspace{-5pt}
\end{table}

A different line of work goes beyond aggregate correctness to study what goes wrong in generated programs. These studies rely on manual inspection to build taxonomies of errors in \gls{LLM}-generated code and identify common failure modes, providing more actionable insights than simple \enquote{pass/fail} metrics~\cite{liu2024exploring, song2023empirical, tambon2025bugs, wang2025towards, chen2024deep, liu2024no, mo2025assessing}. However, existing error analyses exhibit several limitations (Table~\ref{tab:related_work}). Most focus on a single programming language, typically Python, limiting the range of observable errors and generalizability \cite{liu2024exploring, song2023empirical, tambon2025bugs, wang2025towards}. Python’s dynamic execution model masks entire classes of failures related to memory management and undefined behavior, and even extensions to languages such as Java abstract away many system-level error modes common in lower-level languages \cite{chen2024deep}. In addition, prior work often relies on either a small number of problems or a limited number of samples per problem \cite{wang2025towards, song2023empirical, tambon2025bugs, liu2024exploring, liu2024no}, reducing statistical power. Finally, most studies~\cite{liu2024no, mo2025assessing, liu2024exploring, song2023empirical, tambon2025bugs, chen2024deep} evaluate only one or a few models, making it difficult to distinguish model-specific artifacts from general error patterns shared across models and architectures. They also rarely examine how prompting choices affect error distributions, leaving an important dimension of variability underexplored.

Our study analyzes 86,726 runtime or compilation errors across seven models, two prompting techniques (baseline and \gls{CoT} prompting), four compiled programming languages, and 1,651 problems. This scale and diversity let us (i) quantify how error types vary across languages, (ii) separate model-specific quirks from broader trends, (iii) assess the impact of prompting on error behavior, and (iv) measure both compilation and runtime errors with statistically meaningful sample sizes.

\section{Experimental setup}
\label{sec:data}

\begin{figure}[t]
    \centering
    \includegraphics[width=\linewidth, trim=20pt 2pt 20pt 5pt, clip]{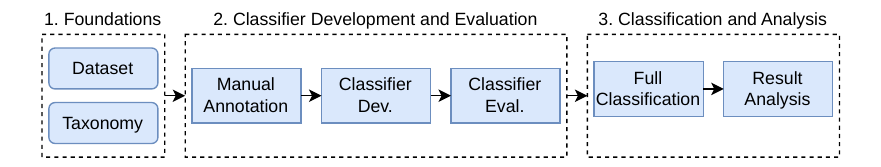}
    \caption{Methodology}
    \label{fig:methodology}
\end{figure}

Our study follows the three-stage methodology shown in Figure~\ref{fig:methodology}. First, foundations are established by selecting a dataset of problem–code–error pairs and an error taxonomy. Then, a subset of the dataset is manually annotated according to the taxonomy and used to assess different \gls{LLM}-based solutions to identify a suitable approach for automated error classification. Finally, in the large-scale classification and analysis stage, the selected solution classifies all samples, and the resulting error distribution is analyzed.

\subsection{Foundations}

The first stage establishes the foundations of the study by defining the dataset of \{code–diagnostic-test-summary\} samples and adopting an existing error taxonomy to guide the subsequent analysis. We intentionally focus on an unassisted setting to capture the natural error tendencies of current models, without external tooling or human intervention.

\subsubsection{Dataset}

The dataset used for this study is based on the PROBE dataset~\cite{nogueira2026probebenchmarkingcodegeneration}. This dataset is made up of 1,651 programming-contest-style problems from IBM's Project CodeNet dataset~\cite{puri2021codenet}, grouped into four difficulty levels. Each problem is accompanied by a set of unit tests and reference solutions and contains a large set of LLM-generated solutions. These solutions were generated by six models (temperature 0.6): four smaller open-source models (Qwen2.5:14b, Qwen2.5-Coder 7B and 14B, and DeepSeek-Coder-v2) and two larger proprietary models (GPT-4.1-mini and Gemini-2.0-Flash). Each model generated five solutions per problem under two prompting strategies (0- and 1-shot) and in four programming languages (C++, Java, C, and Rust), resulting in 396,240 samples. Each sample consists of an initial generation and, if incorrect, up to two additional generations obtained through feedback after testing.

This dataset was chosen due to three key factors. First, it includes a large number of problems spanning multiple difficulty levels, enabling analysis across a wide range of programming challenges. Second, it provides a substantial volume of LLM-generated solutions produced by a diverse set of models, allowing for robust cross-model comparisons of compilation and runtime errors. Finally, the inclusion of a feedback-driven generation process (i.e., iterative re-generation after testing) enables the analysis of how such feedback mechanisms influence error patterns and solution diversity.

To study the impact of prompting, we extend the dataset by generating additional solutions using \gls{CoT} prompting, while keeping all other aspects of the generation process unchanged. In contrast to the original setup, the CoT setting uses single-pass generation without feedback, allowing comparison between baseline and CoT prompting without the confounding effects from feedback-based correction. To provide a more up-to-date picture of model behavior, we further extend both the baseline and CoT settings with GPT-oss:120b, a more recent and larger open-weight model.

The original study reported identical functional correctness, similarity, and code quality metrics for both 0-shot and 1-shot settings across all models and languages. However, the results indicate that, in this context, the inclusion of examples did not impact performance. Therefore, for conciseness, we focus only on the 0-shot and the new \gls{CoT} settings.
We analyze the samples with compilation or runtime errors. Across the two prompting techniques, a total of 52,669 samples failed to compile: 9,953 in C++, 6,996 in Java, 12,304 in C, and 23,416 in Rust (Rust enforces stricter compile-time checks, resulting in more frequent compilation errors). Additionally, 34,057 samples resulted in at least one runtime error during unit testing: 5,605 in C++, 8,279 in Java, 9,135 in C, and 11,038 in Rust.

We focus exclusively on compiled languages (C++, Java, C, and Rust), excluding interpreted languages due to fundamental differences in how errors manifest and are classified. In interpreted settings, distinct failure modes are often collapsed into runtime errors. For example, an undeclared variable may trigger a compilation error in compiled languages, but appear as a runtime error in interpreted ones, despite differing underlying causes. This leads to inconsistencies in error categorization and hinders meaningful cross-language comparisons. Furthermore, we restrict our analysis to compilation and runtime errors, excluding issues such as incorrect outputs or timeouts. Unlike compilation and runtime failures, which provide explicit diagnostic signals, these errors require reasoning about program semantics and intended behavior, making their classification inherently more ambiguous. Focusing on observable error types enables more robust and consistent analysis.

\subsubsection{Taxonomy}

Our goal is not to propose a new taxonomy, but to operationalize and validate existing taxonomies for LLM-generated code in a cross-language, large-scale setting. To this end, we employ two complementary taxonomies to capture distinct failure modes: one for compilation errors and another for runtime errors. Both are based on the schemata proposed by Liu et al.~\cite{liu2024no}, derived from the manual analysis of 728 \gls{LLM}-generated code samples for LeetCode problems.

We chose not to rely on traditional defect taxonomies such as \gls{ODC}, as these are primarily designed for human-written software and operate at a coarser level of abstraction. While such taxonomies can capture underlying defect causes that may lead to compilation failures (e.g., undeclared variables), they do not explicitly distinguish or categorize compilation errors as first-class outcomes. This limitation is particularly relevant in the context of \gls{LLM}-generated code, where compilation errors are both frequent and structurally distinct. \glspl{LLM} often produce error patterns such as missing imports or invalid type annotations, that, in conventional development workflows, would often be caught and corrected immediately through IDE support, compiler feedback, or manual review.

While we adopted the structure proposed by Liu et al.~\cite{liu2024no} as our starting point, several targeted adjustments were made to improve its generality and applicability across multiple languages. Specifically, we refined the taxonomy to remove environment-specific categories tied to the original platform (i.e., LeetCode), merge overlapping definitions, and adjust categories that were either too narrow or ambiguous. In addition, a small number of additional classes were introduced to account for error types not covered in the original schema when extending the taxonomy across languages. The key changes are summarized below.

\begin{table*}[t]
\centering
\renewcommand{\arraystretch}{1}
\caption{Compilation Error Classification}
\label{tab:taxonomy-comp}
\begin{tabular}{p{3.4cm} p{13.9cm}}
\hline
\textbf{Error Type} & \textbf{Description} \\
\hline
Label error & A label can only be part of a statement, and a declaration is not a statement. \\
Redefinition & A symbol has been defined in multiple places. \\
Function declaration error & Function is invoked before declaration. \\
Undeclared variable & Variable is referenced or used in a program without being previously declared or defined. \\
Constant function & Function (or method) is generated with an empty body. \\
Use an undeclared function & Generated code snippet uses undeclared and undefined functions. \\
Incompatible parameter types & Mismatch between the expected parameter type and the actual argument type. \\
Uninitialized variable & Variable is being declared and initialized, but not given a valid initial value before being used. \\
Error of \#include & Indicated file cannot be compiled (e.g., trying to include a file that does not exist). \\
No attribute & Attempt to access a field or method that does not exist in a defined struct/class type. \\
Missing import or \#include & A function or variable from a package is called without the package being imported/included. \\
Missing Public Class & No public class in the code. \\
Ownership and lifetime error & Violations of Rust's ownership, borrowing, or lifetime rules. \\
Trait or type-bound error & Required trait implementations/type constraints not satisfied, or type inference fails due to ambiguous/missing information. \\
Other & Miscellaneous compilation errors not covered by other categories. \\
\hline
\end{tabular}
\end{table*}

\begin{enumerate}[leftmargin=*]
    
    \item \textbf{LeetCode-specific labels removed.} Labels referencing LeetCode artifacts (e.g., method name templates) do not apply to our setting and were removed.
  
    \item \textbf{\enquote{Syntax Error} removed.} This broad class overlapped with more specific categories such as \enquote{Label error} and \enquote{Redefinition} and was removed to avoid redundancy.

    \item \textbf{\enquote{Invalid operators} and \enquote{Incompatible parameter types} merged.} Both capture type-related mismatches, one in expressions, and the other in function arguments. These were consolidated into a single \enquote{Incompatible parameter types} category to improve clarity and cohesion.
  
    \item \textbf{\enquote{Missing delimiters} removed.} This fine-grained syntactic class (e.g., missing commas or brackets) operates at a lower level of abstraction than the remaining categories and was removed to maintain consistency.
  
    \item \textbf{\enquote{Missing imports} category added.} Omitting required imports was a recurring error mode in our corpus not covered by the original taxonomy.
  
    \item \textbf{Java-specific \enquote{Missing Public Class} added.} Unlike LeetCode, standalone Java files require a public class to compile; we added an explicit category to capture this.
  
    \item \textbf{Rust-specific errors added.} Two categories for Rust-specific compilation errors were introduced: (i) \enquote{Ownership and lifetime errors} for violations of Rust's ownership and borrowing rules, and \enquote{Trait or type-bound errors} for unsatisfied trait constraints or failed type inference.
    
    \item \textbf{\enquote{Other} bucket added.} Covers infrequent or unexpected compilation errors not fitting any defined category.
\end{enumerate}

For the Runtime Errors, the following changes were made:

\begin{enumerate}[leftmargin=*]
    \item \textbf{\enquote{Heap-buffer-overflow} category replaced by broader \enquote{Out-of-memory} definition.} 
    The original category overlapped conceptually with \enquote{Out-of-bound} and did not generalize to managed languages.  
    To improve clarity and cross-language applicability, we redefined \enquote{Out-of-memory} to cover all memory allocation failures and extended \enquote{Out-of-bound} to all invalid indexing operations.
    \item \textbf{\enquote{Other} bucket added.} Covers infrequent or unexpected runtime errors not fitting any defined category.
\end{enumerate}

The resulting taxonomies are designed to support consistent error classification across multiple programming languages while preserving language-specific distinctions where necessary (e.g., ownership errors in Rust or missing public classes in Java). This enables a systematic, large-scale empirical analysis of LLM-generated code errors across languages, rather than the introduction of a new taxonomy. The complete taxonomies are shown in Tables~\ref{tab:taxonomy-comp} and~\ref{tab:taxonomy-runtime}.

\subsection{Classifier Development and Evaluation}

The goal of this stage is to design a solution for mapping each \{code + diagnostic + test-summary\} sample to taxonomy labels. While manual classification is preferable for quality, it is infeasible at the scale of our dataset; therefore, we build and evaluate automated, LLM-based classification solutions on a human-labeled subset and then select the best-performing solution for large-scale application.

\subsubsection{Manual annotation}
\label{sec:annotation}

A labeled evaluation set is required both to measure model performance and to validate the taxonomies. For each error family (compilation and runtime), we randomly sampled 300 examples, stratified by language (75 per target language). Each sampled example consists of the generated code, the associated error message, and, for runtime errors, the unit-test input that triggered the error. \rn{Although this approach does not ensure coverage of all taxonomy classes, it preserves the original error distribution within each language.} The following annotation protocol was followed:

\begin{enumerate}

    \item \textbf{Independent labeling.} Two experts independently label each example according to the taxonomy.
    
    \item \textbf{Agreement.} Labels are compared and inter-annotator agreement is measured (see metrics below). In case of disagreement, reviewers presented their points of view and discussed until an agreement was reached.
    
\end{enumerate}

The inter-annotator agreement was measured using the Jaccard index for compilation errors (as the categories are not mutually exclusive) and Cohen's $\kappa$ for runtime errors (as the categories are mutually exclusive). For compilation errors, a Jaccard index of around 0.8 was obtained, with the observed differences between annotators mainly stemming from a small number of recurring errors that were interpreted differently. For runtime errors, Cohen's $\kappa$ was around 0.62, indicating substantial agreement~\cite{kappa}, especially considering the wide variety of classes and the fact that many samples were inherently complex to annotate.

\begin{table*}[t]
\centering
\renewcommand{\arraystretch}{1}
\caption{Runtime Error Classification}
\label{tab:taxonomy-runtime}
\begin{tabular}{p{3.5cm} p{13.8cm}}
\hline
\textbf{Error Type} & \textbf{Description} \\
\hline
1. Numeric overflow & The result of an arithmetic operation exceeds the maximum value that can be represented by a given numeric data type. \\
2. Out of memory & The program fails to allocate memory due to insufficient heap or system resources. \\
3. Undefined behavior & The execution of a certain code statement is not defined or predictable, potentially leading to unexpected runtime errors or crashes (e.g., left shift of negative value). \\
4. Out of bounds & Accessing a data structure at an index or memory address that lies outside its valid range. \\
5. Constant function & Function (or method) is generated with an empty body. \\
6. Null pointer dereference & Accessing a memory location or object that is explicitly null due to a programming logic error. \\
7. Type error & An operation is performed on an object of an incompatible type. \\
8. Value error & A method or operation receives a valid type of input. Still, the value itself is not suitable or within the expected range for the specific operation (e.g., calling \texttt{.max()} on an empty array in Java). \\
9. Heap use-after-free & Use a memory block in the heap after it has been deallocated (freed). \\
10. Recursion error & A function/method calls itself recursively without reaching a base case or termination condition. \\
11. Uninitialized variable & Variable is being declared and initialized, but not given a valid initial value before being used. \\
12. Undeclared variable & Variable is referenced or used in a program without being previously declared or defined. \\
13. No attribute & The accessed member does not exist in the corresponding data structure. \\
14. Divided by zero & Attempt to divide a number by zero. \\
15. Incorrect input processing & Errors caused by wrong assumptions about the input format. \\
-1. Other & Miscellaneous runtime errors not covered by other categories. \\
\hline
\end{tabular}
\end{table*}

\subsubsection{Classifier Development}

We design a zero-shot \gls{LLM}-based classifier to map \{code + diagnostic + test-summary\} samples to taxonomy labels, enabling scalable analysis of compilation and runtime errors across a large corpus of generated programs. The emphasis at this stage is on identifying a sufficiently reliable and stable configuration for automated labeling, rather than on conducting a comparative evaluation of \glspl{LLM}. We evaluate \textbf{GPT-4.1-mini} with direct and \gls{CoT} prompting, and \textbf{Gemini-2.0-Flash} with direct prompting only. \rn{Gemini-2.0-Flash was not evaluated with \gls{CoT} because it already performed substantially worse than GPT-4.1-mini under direct prompting. Given that gap, \gls{CoT} was unlikely to make it the best-performing classifier, and the additional inference cost was not warranted.} All models are evaluated with temperature 0, top-$k$ of 50, and top-$p$ of 1.0 to promote deterministic and reproducible outputs, and are accessed via their respective APIs (OpenAI and Google Vertex AI, respectively). While strict determinism is not guaranteed for commercial \glspl{LLM}, this configuration produced consistent outputs across runs.

\subsubsection{Classifier Evaluation}

\setlength{\tabcolsep}{2.2pt} 
\begin{table}[t]
\vspace{6pt}
\centering
\caption{Exact match accuracy of the classifiers}
\vspace{-4pt}
\label{tab:model-perf}
\begin{tabular}{c|cccc|cccc}
\multirow{2}{*}{Model} & \multicolumn{4}{c|}{Compilation}                              & \multicolumn{4}{c}{Runtime}                                   \\
        & C++  & Java       & C        & Rust       & C++       & Java          & C      & Rust          \\ \hline
GPT-4.1-mini & 0.91 & \textbf{0.91} & \textbf{0.84} & \textbf{0.91} & \textbf{0.81} & 0.91 & 0.83 & \textbf{0.84} \\
Gemini-2.0-Flash & 0.80 & 0.88 & 0.59 & 0.79 & 0.71 & 0.80 & 0.53 & 0.72 \\
GPT-4.1-mini (CoT) & \textbf{0.92} & 0.85 & 0.83 & 0.81 & 0.80 & \textbf{0.92} & \textbf{0.84} & 0.83         
\end{tabular}
\end{table}

Classifiers are evaluated against the manually annotated dataset to identify the best-performing model. Evaluation is based on strict accuracy: a prediction is considered correct only if the set of errors exactly matches the human annotations. We do not report macro-averaged precision or recall, as these metrics proved unstable due to the rarity of several error classes. Results for compilation and runtime error classification are summarized in Table~\ref{tab:model-perf}.

For compilation errors, GPT-4.1-mini achieved the best overall performance, with accuracies of 0.91 for C++, 0.91 for Java, 0.84 for C, and 0.91 for Rust without \gls{CoT}. The highest accuracy (0.92) was achieved with C++ and \gls{CoT}. Gemini-2.0-Flash consistently underperformed, particularly for C (0.59). For runtime errors, GPT-4.1-mini again outperformed Gemini-2.0-Flash across all languages. Without \gls{CoT}, accuracies ranged from 0.81 to 0.91. \gls{CoT} prompting yielded marginal improvements for Java and C, but slightly reduced performance for C++ and Rust. Given the negligible gains and additional inference cost, GPT-4.1-mini (no \gls{CoT}) was selected for both compilation and runtime error classification.

Table~\ref{tab:detailed_performance} reports per-class recall and precision (with support) for the selected model (GPT-4.1-mini) on compilation and runtime errors. Only classes with at least three samples in at least one language are included. Results show strong performance on common error classes, while rarer ones exhibit greater variability due to limited representation. The model also shows lower recall for the \enquote{Other} category, indicating difficulty identifying this residual class. Several classes are not observed for certain languages. As the dataset was built by randomly sampling 75 errors per language, their absence suggests they are rare and have limited impact on aggregate trends. For runtime errors, results indicate a bias toward predicting \enquote{Out-of-Bound} errors, reducing recall for classes such as \enquote{Out-of-memory} and \enquote{Recursion error}. Manual analysis revealing that this was due to the model having a tendency to default to an \enquote{Out-of-Bound} classification when a segmentation fault is observed, and thus error information is limited.

Overall, while strict accuracy ranges between 0.81 and 0.92, the per-class analysis shows that remaining errors are concentrated among a small subset of classes and follow consistent patterns.  \rn{To quantify the impact of these errors on the final results, we applied a confusion-matrix-based correction: for each class, the observed corpus count is scaled by the ratio of precision to recall in the manually annotated evaluation set, yielding a bias-corrected proportion. Although the full per-class breakdown is omitted here for space, this correction did not alter any of the overall trends reported in Section~\ref{sec:results}.}

\begin{table*}[t]
\centering
\caption{Recall/Precision and support for GPT-4.1-mini across Compilation (above) and Runtime (below) errors}
\label{tab:detailed_performance}
\begin{subtable}{\textwidth}
\centering
\begin{tabular}{l|lllllllll}
Category & -1 & 3 & 4 & 6 & 7 & 13 & 14 & 15 & 16 \\
\midrule
C++ & 0.50/0.67 (4) & 0.50/1.00 (2) & 1.00/0.91 (10) & 1.00/0.25 (1) & 0.89/0.94 (19) & 0.95/0.98 (44) & 0.00/0.00 (0) & 0.00/0.00 (0) & 0.00/0.00 (0) \\
Java & 0.69/1.00 (13) & 0.00/0.00 (0) & 0.88/1.00 (17) & 1.00/1.00 (1) & 0.96/1.00 (28) & 1.00/1.00 (8) & 1.00/0.75 (6) & 0.00/0.00 (0) & 0.00/0.00 (0) \\
C & 0.25/0.50 (4) & 0.60/1.00 (5) & 0.89/1.00 (9) & 0.90/0.75 (10) & 0.96/0.87 (27) & 1.00/0.96 (25) & 0.00/0.00 (0) & 0.00/0.00 (0) & 0.00/0.00 (0) \\
Rust & 0.25/1.00 (4) & 0.00/0.00 (1) & 1.00/1.00 (1) & 1.00/0.50 (1) & 0.95/0.97 (37) & 1.00/1.00 (18) & 0.00/0.00 (0) & 1.00/0.83 (19) & 0.83/0.83 (12) \\
\bottomrule
\end{tabular}
\end{subtable}

\vspace{0.8em}

\begin{subtable}{\textwidth}
\centering
\begin{tabular}{l|llllllll}
Category & -1 & 1 & 2 & 3 & 4 & 10 & 14 & 15 \\
\midrule
C++ & 0.00/0.00 (1) & 0.80/1.00 (5) & 0.67/0.67 (6) & 0.80/0.80 (5) & 0.97/0.78 (33) & 0.69/1.00 (13) & 1.00/1.00 (5) & 0.25/0.50 (4) \\
Java & 0.00/0.00 (0) & 0.38/1.00 (8) & 1.00/0.50 (1) & 0.00/0.00 (0) & 1.00/0.97 (31) & 0.92/1.00 (12) & 1.00/1.00 (1) & 0.95/0.83 (21) \\
C & 0.00/0.00 (3) & 1.00/1.00 (2) & 0.81/0.88 (26) & 1.00/1.00 (2) & 0.94/0.83 (32) & 0.67/0.67 (3) & 1.00/1.00 (3) & 0.00/0.00 (0) \\
Rust & 0.00/0.00 (3) & 1.00/1.00 (29) & 0.00/0.00 (0) & 0.00/0.00 (0) & 1.00/0.64 (14) & 1.00/1.00 (1) & 0.00/0.00 (0) & 0.68/0.95 (28) \\
\bottomrule
\end{tabular}
\end{subtable}
\end{table*}

\subsection{Classification and analysis}
\label{sec:classification_analysis}

Having selected an LLM-based solution to map generated samples to taxonomy labels, we apply it to the entire dataset and conduct a comprehensive analysis of model error behavior.

\subsubsection{Full classification}
\label{sec:full_classification}

The classifier is run over all samples.
 
\subsubsection{Analysis}
\label{sec:analysis}

\setlength{\tabcolsep}{3pt} 
\begin{table*}[h]
    \centering
    \caption{Compilation Error Distribution per language (left) and problem difficulty (right) (baseline prompt)}
    \label{tab:comp-lang}
    \begin{tabular}{lllll|rrrr}
    \toprule
     & C++ & Java & C & Rust & Diff. 0 & Diff. 1 & Diff. 2 & Diff. 3 \\
    \midrule
    Undeclared var. & \textbf{410 (6.1\%)} & \textbf{566 (15.7\%)} & \textbf{713 (10.4\%)} & \textbf{490 (3.8\%)} & \textbf{330 (4.4\%)} & \textbf{854 (7.6\%)} & \textbf{777 (9.1\%)} & \textbf{218 (7.9\%)}  \\
    Use undec. func. & 454 (6.8\%) & 64 (1.8\%) & 1150 (16.8\%) & 177 (1.4\%) & 447 (5.9\%) & 774 (6.8\%) & 461 (5.4\%) & 163 (5.9\%) \\
    Inc. param. types & \textbf{1358 (20.3\%)} & \textbf{1551 (43.0\%)} & \textbf{1497 (21.9\%)} & \textbf{5616 (43.4\%)} & \textbf{2209 (29.4\%)} & \textbf{3916 (34.6\%)} & \textbf{2954 (34.7\%)} & \textbf{943 (34.0\%)} \\
    Uninitialized var. & 25 (0.4\%) & 63 (1.7\%) & 333 (4.9\%) & 79 (0.6\%) & 136 (1.8\%) & 214 (1.9\%) & 103 (1.2\%) & 47 (1.7\%) \\
    No attribute & 238 (3.5\%) & 179 (5.0\%) & 188 (2.7\%) & 438 (3.4\%) & 224 (3.0\%) & 384 (3.4\%) & 322 (3.8\%) & 113 (4.1\%) \\
    Missing import & \textbf{3795 (56.6\%)} & \textbf{292 (8.1\%)} & \textbf{2432 (35.5\%)} & \textbf{2656 (20.5\%)} & \textbf{2924 (38.9\%)} & \textbf{3318 (29.3\%)} & \textbf{2219 (26.1\%)} & \textbf{714 (25.7\%)} \\
    Miss. Pub. Class & 0 (0.0\%) & 593 (16.4\%) & 0 (0.0\%) & 1 (0.0\%) & 116 (1.5\%) & 181 (1.6\%) & 234 (2.8\%) & 63 (2.3\%) \\
    Own. and life. & 2 (0.0\%) & 0 (0.0\%) & 0 (0.0\%) & 2165 (16.7\%) & 618 (8.2\%) & 753 (6.7\%) & 592 (7.0\%) & 204 (7.4\%) \\
    Trait/type bound & 0 (0.0\%) & 2 (0.1\%) & 0 (0.0\%) & 784 (6.1\%) & 190 (2.5\%) & 295 (2.6\%) & 226 (2.7\%) & 75 (2.7\%) \\
    Other & 107 (1.6\%) & 129 (3.6\%) & 201 (2.9\%) & 334 (2.6\%) & 128 (1.7\%) & 255 (2.3\%) & 279 (3.3\%) & 109 (3.9\%) \\ \hline
    \textbf{Total} & 6706 (100\%) & 3611 (100\%) & 6850 (100\%) & 12942 (100\%) & 7515 (100\%) & 11311 (100\%) & 8508 (100\%) & 2775 (100\%) \\
    \end{tabular}
\end{table*}

The results are analyzed in detail. We start by examining the overall distribution of error types for baseline prompting and how it varies with factors such as programming language, model, and problem difficulty. Next, we study how iterative feedback influences each error type. Finally, we analyze the impact of \gls{CoT} prompting on the error distribution.

To assess whether the observed differences in error frequencies are statistically significant, we use two-proportion z-tests. We chose z-tests because our sample sizes are large, making the normal approximation appropriate, and because the test is simple, standard, and directly interpretable. 


\section{Results and Discussion}\label{sec:results}

In this section, we present a comprehensive analysis of the classification results, aiming to characterize the types and distributions of errors produced by the models. We first examine compilation errors, followed by runtime errors. For each category, we analyze its distribution and explore how programming language, model, problem difficulty, and prompting influence these patterns. 
Due to space constraints, all tables in this section display only the results for labels with a frequency of above three percent in any column (i.e., any language for the language tables and any model for the model tables).

\subsection{Compilation Errors}

Table~\ref{tab:comp-lang} presents the distribution of compilation errors per programming language (left) and per problem difficulty (right) for the baseline prompt. Each cell reports both the number of errors in both absolute and relative terms (within that language or difficulty level). Two dominant error types account for most compilation errors across all languages: \enquote{Missing import} and \enquote{Incompatible parameter types}, followed by \enquote{Undeclared variable}. Rust shows a distinct pattern, with 16.7\% of errors related to \enquote{Ownership and lifetime} and 6.1\% to \enquote{Trait/type bound} issues. Examples of the two most common compilation error types are shown in Listings~\ref{lst:missing-import} and~\ref{lst:incompatible-types}. The first example is a code sample generated by GPT-4.1-mini, which fails to compile because \texttt{uint32\_t} is used without including \texttt{<cstdint>}. The second example illustrates an \enquote{Incompatible parameter types} error in Java, where a primitive variable is combined with a \texttt{BigInteger} using the modulus operator, leading to a type mismatch.

\begin{figure} 
    \begin{lstlisting}[style=smallcpp, label={lst:missing-import}, caption={Example of a missing import error}]
#include <iostream> 
#include <bitset> 
int main() { 
    uint32_t a, b; 
    ... 
    \end{lstlisting}
\vspace{-7pt}
\end{figure}

\begin{figure}[b]
    \begin{lstlisting}[style=smalljava, label={lst:incompatible-types}, caption={Example of a incomp. param. types error}]
public class BenchPhotoProbability {
    private static final BigInteger MOD = new BigInteger(""1000000007"");
    public static void main(String[] args) {
        ...
        long p = 0, q = 0, r = 0;
        ...
        for (int i = 0; i < N; ++i) {
            if (s.charAt(i) == 'X') {
                ...
                p %= MOD;
                q %= MOD;
                r %= MOD;
                ...
    \end{lstlisting}
\vspace{-12pt}
\end{figure}

Two additional insights can be highlighted. First, the total number of errors labeled as \enquote{Other} is low, indicating that the taxonomy captures the vast majority of issues. Although it is unlikely the model captured all errors that should belong to this category, the recall and accuracy for this class suggest that the real number is still small. Additionally, when examining the language-specific error classes (\enquote{Missing Public class} for Java and \enquote{Ownership and lifetime error} and \enquote{Trait or type bound error} for Rust), we observe that the models correctly attribute these classes only to examples of the respective language. Across the dataset, \enquote{Missing Public Class} was assigned 594 times in total, 593 of which (99.8\%) were in Java. Likewise, the combined total for \enquote{Ownership and lifetime error} and \enquote{Trait or type bound error} is 2,953 classifications, of which 2,949 (99.9\%) belong to Rust. These results indicate that the models correctly map language-specific compiler diagnostics to the corresponding taxonomy labels.

When we break down the distribution per language, some clear trends emerge. For C and C++, \enquote{Missing import} is by far the most common classification, accounting for 56.6\% of C++ compilation errors and 35.5\% of C compilation errors. The same cannot be said for the other two languages, as \enquote{Missing import} accounts for only 8.1\% of compilation errors in Java and 20.5\% in Rust. All pairwise differences between languages are statistically significant (two-proportion z-tests with Bonferroni correction and $\alpha$=0.05).
These differences can be explained by how standard libraries are designed across these languages. C and C++ require explicit includes for nearly all functionality (even basic operations like I/O need \#include $<$stdio.h$>$ or \#include $<$iostream$>$). Java and Rust, however, take a different approach: Java automatically imports Java.lang.*, while Rust's prelude brings common types and traits into scope by default. As a result, \glspl{LLM} generating Java or Rust code is less likely to omit necessary imports because fewer explicit import statements are needed to begin with.

\begin{table*}[t]
    \centering
    \caption{Compilation error distribution per model (C++) (baseline prompt)}
    \label{tab:comp-model-cpp}
    \begin{tabular}{llllllll}
    \toprule
     & deepseek-coder-v2:16b & gemini-2.0-flash & gpt-4.1-mini & gpt-oss:120b & qwen2.5-coder:14b & qwen2.5-coder:7b & qwen2.5:14b \\
    \midrule
    Undeclared variable & \textbf{95 (9.3\%)} & \textbf{3 (0.5\%)} & \textbf{11 (3.2\%)} & \textbf{1 (1.0\%)} & \textbf{40 (4.1\%)} & \textbf{100 (8.3\%)} & \textbf{160 (6.4\%)} \\
    Use undeclared function & 80 (7.9\%) & 11 (1.9\%) & 17 (4.9\%) & 3 (2.9\%) & 74 (7.6\%) & 70 (5.8\%) & 199 (8.0\%) \\
    Incompatible param. types & \textbf{198 (19.4\%)} & \textbf{15 (2.6\%)} & \textbf{49 (14.1\%)} & \textbf{30 (29.4\%)} & \textbf{227 (23.4\%)} & \textbf{270 (22.3\%)} & \textbf{569 (22.9\%)} \\
    Uninitialized variable & 3 (0.3\%) & 0 (0.0\%) & 3 (0.9\%) & 0 (0.0\%) & 4 (0.4\%) & 4 (0.3\%) & 11 (0.4\%) \\
    No attribute & 29 (2.8\%) & 1 (0.2\%) & 9 (2.6\%) & 8 (7.8\%) & 34 (3.5\%) & 71 (5.9\%) & 86 (3.5\%) \\
    Missing import & \textbf{571 (56.0\%)} & \textbf{532 (93.2\%)} & \textbf{198 (57.1\%)} & \textbf{1 (1.0\%)} & \textbf{542 (55.9\%)} & \textbf{620 (51.2\%)} & \textbf{1331 (53.5\%)} \\
    Other & 12 (1.2\%) & 3 (0.5\%) & 14 (4.0\%) & 15 (14.7\%) & 10 (1.0\%) & 21 (1.7\%) & 32 (1.3\%) \\ \hline
    \textbf{Total} & 1019 (100\%) & 571 (100\%) & 347 (100\%) & 102 (100\%) & 970 (100\%) & 1211 (100\%) & 2486 (100\%) \\
    \end{tabular}
\end{table*}

The \enquote{Incompatible parameter types} error was more prevalent in Java (43.0\%) and Rust (43.4\%) than in C++ (20.3\%) and C (21.9\%). The differences are statistically significant between all models pairs except between C++ and C, and between Java and Rust (two-proportion z-tests with Bonferroni correction and $\alpha$=0.05).
In Java, this higher proportion is primarily a relative effect: the absolute number of errors in this category (1,551) is comparable to that in C++ (1,358) and C (1497), so the percentage difference primarily reflects the much lower number of missing import errors in Java. 
In contrast, Rust exhibits a substantially higher absolute number of incompatible parameter type errors (5,555), indicating a genuine increase rather than a relative effect. This observation is consistent with prior work showing that \glspl{LLM} struggle with Rust in code-related tasks such as code translation~\cite{vieira2025Polyglot}. While the tasks differ, both findings suggest that Rust poses particular challenges for \glspl{LLM}, potentially due to its strict type system and lower representation in training data.

When focusing on the per-problem difficulty distribution, we observe that, although the overall shifts in error distributions are not large, the distribution varies significantly across difficulty levels (chi-square tests with Bonferroni correction and $\alpha$=0.05).
Some notable trends can be observed. The relative frequency of Missing imports was higher in easier problems, dropping from 38.9\% in difficulty 0 to 25.7\% in difficulty 3. Conversely, incompatible parameter types and use of undeclared functions showed the opposite pattern, with incompatible parameter types increasing from 29.4\% to 34.7\% between difficulty levels 0 and 2, while undeclared variables rose from 4.4\% to 9.1\%. This is due to more complex problems involving more variables and function calls, increasing the likelihood of parameter mismatches or undeclared references. In simpler problems, where fewer functions and variables are needed, these errors are less common, allowing other error types, such as missing imports, to dominate the relative frequency. In simpler problems, the relative scarcity of these error types inflates the share of missing imports.

Table~\ref{tab:comp-model-cpp} presents the per-model distribution of C++ errors (baseline prompt). We present this analysis for a single language to ensure observed differences are not biased by language-specific error profiles (e.g., \enquote{Missing import} errors are more frequent in C/C++).

We observe that error types vary significantly across models. \enquote{Missing import} errors are the most frequent category across all models except GPT-oss:120b, which makes only a single such error. There are also differences among the remaining models, with missing imports being particularly dominant for Gemini-2.0-Flash (93.2\%). Using two-proportion z-tests over all model pairs with Bonferroni correction and $\alpha$=0.05, GPT-oss:120b and Gemini-2.0-Flash differ significantly from all other models, while most other pairwise differences are not statistically significant. Besides that, smaller models tend to make more \enquote{Use undeclared function} and \enquote{Incompatible parameter types} errors, both in relative and absolute terms, except for GPT-oss:120b, whose high relative share of \enquote{Incompatible parameter types} (29.4\%) is not mirrored in absolute terms, as it makes very few errors overall due to rarely making other errors such as missing imports. These differences may stem from variations in models' training data (e.g., including more LeetCode submissions that lack includes) or training procedures. We also observe that qwen2.5-coder:14b produces substantially fewer compilation errors than its general-purpose counterpart, while maintaining a similar distribution of error types. This suggests that code-specific fine-tuning primarily reduces the frequency of errors rather than altering their nature. Although analyses for other languages were excluded due to space constraints, these also revealed apparent differences in error distributions across models.

Finally, Table~\ref{tab:feedback-comp} shows the impact of feedback on compilation errors. An error is \emph{fixed} if the code passes all unit tests, \emph{partially fixed} if the error is removed but tests still fail, and \emph{not fixed} if it remains after feedback. We can see that six out of the ten error types, are fixed completely between 12.4\% and 15.4\% of cases. In contrast, \enquote{Undeclared variable} (5.3\%), \enquote{Incompatible parameter types} (7.8\%), and \enquote{Use undeclared function} (7.9\%) are fixed completely far less often. This gap can be explained by the nature of the fix required: the first group is typically resolved with a single, local edit (e.g., adding an import or initializing a variable), whereas the latter group requires reasoning about how a variable or parameter is used elsewhere in the code, making a complete fix harder even once the error is identified.

\begin{table}[b]
    \vspace{12pt}
    \centering
    \caption{Impact of feedback on the compilation errors}
    \label{tab:feedback-comp}
    \begin{tabular}{llll}
    \toprule
    Error & Not Fixed & Partially fixed & Fixed \\
    \midrule
    Undeclared variable & \textbf{23.6\%} & \textbf{71.0\%} & \textbf{5.3\%} \\
    Use undeclared function & 13.7\% & 78.4\% & 7.9\% \\
    Incompatible param. types & \textbf{33.7\%} & \textbf{58.5\%} & \textbf{7.8\%} \\
    Uninitialized variable & 21.8\% & 62.8\% & 15.4\% \\
    No attribute & 15.7\% & 70.0\% & 14.3\% \\
    Missing import & \textbf{19.6\%} & \textbf{68.0\%} & \textbf{12.4\%} \\
    Missing Public Class & 2.2\% & 83.7\% & 14.1\% \\
    Owner. and life. err. & 33.6\% & 53.5\% & 12.9\% \\
    Trait or type bound error & 18.1\% & 69.0\% & 13.0\% \\
    Other & 12.2\% & 78.2\% & 9.6\% \\
    \bottomrule
    \end{tabular}
\vspace{-7pt}
\end{table}

Table~\ref{tab:CoT-comp} shows the impact of \gls{CoT} prompting on compilation errors for C++. We present results for a single language to isolate the effect of prompting from language-specific differences, following the same rationale as in the per-model analysis. Analyses for the remaining languages lead to consistent conclusions. Overall, the distribution shifts moderately, but some clear patterns emerge. Most notably, \enquote{Missing import} errors become substantially less prevalent (-15.9\%), increasing the relative share of all other error types.

\begin{table}[t]
    \centering
    \caption{Impact of CoT on Compilation Errors (C++)}
    \label{tab:CoT-comp}
    \begin{tabular}{ll}
    \toprule
    Undeclared variable & 6.8\% (+0.7\%) \\
    Use undeclared function & 12.5\% (+5.7\%) \\
    Incompatible parameter types & 23.7\% (+3.4\%) \\
    No attribute & 6.3\% (+2.8\%) \\
    Missing import & 40.7\% (-15.9\%) \\
    Other & 4.2\% (+2.6\%) \\
    \hline
    Total & 5909 (-797) \\
    \end{tabular}
\end{table}

\subsection{Runtime Error}

\setlength{\tabcolsep}{3pt} 
\begin{table*}[h]
    \centering
    \caption{Runtime Error Distribution per language (left) and problem difficulty (right) (baseline prompt)}
    \label{tab:runtime-lang}
    \begin{tabular}{lllll|rrrr}
    \toprule
     & C++ & Java & C & Rust & Diff. 0 & Diff. 1 & Diff. 2 & Diff. 3 \\
    \midrule
    Numeric-overflow & 62 (3.8\%) & 220 (5.4\%) & 145 (4.9\%) & 2508 (44.0\%) & 997 (25.6\%) & 1153 (20.4\%) & 559 (15.0\%) & 226 (21.3\%) \\
    Out-of-memory & 173 (10.7\%) & 62 (1.5\%) & 946 (32.0\%) & 30 (0.5\%) & 175 (4.5\%) & 672 (11.9\%) & 311 (8.3\%) & 53 (5.0\%) \\
    Undefined-behavior & 53 (3.3\%) & 4 (0.1\%) & 72 (2.4\%) & 25 (0.4\%) & 17 (0.4\%) & 72 (1.3\%) & 53 (1.4\%) & 12 (1.1\%) \\
    Out-of-bound & \textbf{970 (60.0\%)} & \textbf{1845 (45.5\%)} & \textbf{1510 (51.1\%)} & \textbf{1603 (28.1\%)} & \textbf{1405 (36.1\%)} & \textbf{2256 (40.0\%)} & \textbf{1755 (47.0\%)} & \textbf{512 (48.3\%)} \\
    Null pointer dereference & 42 (2.6\%) & 140 (3.5\%) & 65 (2.2\%) & 2 (0.0\%) & 38 (1.0\%) & 99 (1.8\%) & 80 (2.1\%) & 32 (3.0\%) \\
    Recursion error & \textbf{130 (8.0\%)} & \textbf{580 (14.3\%)} & \textbf{35 (1.2\%)} & \textbf{166 (2.9\%)} & \textbf{83 (2.1\%)} & \textbf{343 (6.1\%)} & \textbf{405 (10.8\%)} & \textbf{80 (7.6\%)} \\
    Divided by zero & 75 (4.6\%) & 95 (2.3\%) & 92 (3.1\%) & 41 (0.7\%) & 104 (2.7\%) & 126 (2.2\%) & 68 (1.8\%) & 5 (0.5\%) \\
    Incorrect input processing & \textbf{69 (4.3\%)} & \textbf{1045 (25.8\%)} & \textbf{36 (1.2\%)} & \textbf{1264 (22.2\%)} & \textbf{1004 (25.8\%)} & \textbf{842 (14.9\%)} & \textbf{453 (12.1\%)} & \textbf{115 (10.9\%)} \\
    \hline
    \textbf{Total} & 1616 (100\%) & 4051 (100\%) & 2955 (100\%) & 5700 (100\%) & 3888 (100\%) & 5642 (100\%) & 3733 (100\%) & 1059 (100\%) \\
    \end{tabular}
\end{table*}

Table~\ref{tab:runtime-lang} presents the per-language (left) and per-problem difficulty (right) distribution of runtime errors as classified by the taxonomy. As in the compilation errors, we can see that only a very small percentage of errors are classified as \enquote{Other}, which is a strong indication that our taxonomy covers most of the runtime errors commonly present in \gls{LLM}-generated code.

Starting with the per-language distribution, it is evident that, as observed with compilation errors, the types of runtime errors produced by the models show clear language-dependent patterns. The most common categories across all languages are \enquote{Out-of-bound} errors, followed by \enquote{Incorrect input processing}, \enquote{Out-of-memory} errors, and \enquote{Recursion} errors. \enquote{Out-of-memory} errors are particularly prevalent in C++ (10.3\%) and C (33.5\%), while they are rare in Java (2.2\%) and completely absent in Rust. These differences are statistically significant for all pairwise comparisons (two-proportion z-tests with Bonferroni correction and $\alpha$=0.05). This discrepancy can be attributed to differences in memory management. In C and C++, memory allocation and deallocation are handled manually, and errors such as allocation of excessively large arrays can easily result in runtime errors. In contrast, Java and Rust employ automatic memory management, Java through garbage collection and Rust through its ownership and borrowing system, substantially reducing the likelihood of explicit out-of-memory crashes during execution. Between C and C++, the observed difference can mostly be attributed to the richer standard library and higher-level abstractions available in C++, which enable flexible and incremental memory usage, whereas C exposes only low-level memory primitives, making programs more sensitive to allocation size decisions and out-of-memory failures.

\enquote{Out-of-bound} errors are very common across all languages. Although the percentage is higher in C++ (59.9\%), C (49.9\%), and Java (45.7\%) than in Rust (28.2\%), this lower percentage does not reflect fewer absolute occurrences. Rather, this is caused by a higher number of errors of other types. Listing~\ref{lst:oob-java} shows a Java example. The dynamic programming array $dp$ is indexed using $i - 2$ inside the loop without guarding against small values of $i$, meaning that, when $i = 1$, the program attempts to access $dp[-1]$, resulting in an error.

\begin{figure}[h]
\vspace{-7pt}
\begin{lstlisting}[style=smalljava, caption={Example of an out-of-bounds error.}, label={lst:oob-java}]
int N = scanner.nextInt();
long[][] dp = new long[N + 1][3];
for (int i = 0; i <= N; i++) {
    if (i == 2) { ...
    } else if (i != 0) {
        dp[i][0] = (dp[i - 2][0] + dp[i - 2][1] + dp[i - 2][2])
        ...
\end{lstlisting}
\vspace{-12pt}
\end{figure}

\begin{table*}[t]
    \centering
    \caption{Runtime error distribution per model (Java) (baseline prompt)}
    \label{tab:runtime-model-cpp}
    \begin{tabular}{llllllll}
    \toprule
    & deepseek-coder-v2:16b & gemini-2.0-flash & gpt-4.1-mini & gpt-oss:120b & qwen2.5-coder:14b & qwen2.5-coder:7b & qwen2.5:14b \\
    \midrule
    Numeric-overflow & 74 (9.6\%) & 28 (6.7\%) & 10 (2.9\%) & 0 (0.0\%) & 31 (5.7\%) & 46 (5.0\%) & 31 (3.4\%) \\
    Out-of-memory & 20 (2.6\%) & 14 (3.4\%) & 0 (0.0\%) & 2 (1.5\%) & 9 (1.6\%) & 10 (1.1\%) & 7 (0.8\%) \\
    Out-of-bound & \textbf{343 (44.4\%)} & \textbf{168 (40.3\%)} & \textbf{136 (39.4\%)} & \textbf{55 (41.7\%)} & \textbf{248 (45.3\%)} & \textbf{435 (47.6\%)} & \textbf{460 (49.8\%)} \\
    Null pointer dereference & 16 (2.1\%) & 10 (2.4\%) & 7 (2.0\%) & 16 (12.1\%) & 19 (3.5\%) & 22 (2.4\%) & 50 (5.4\%) \\
    Recursion error & 111 (14.4\%) & 118 (28.3\%) & 103 (29.9\%) & 28 (21.2\%) & 63 (11.5\%) & 72 (7.9\%) & 85 (9.2\%) \\
    Divided by zero & 26 (3.4\%) & 5 (1.2\%) & 9 (2.6\%) & 1 (0.8\%) & 14 (2.6\%) & 15 (1.6\%) & 25 (2.7\%) \\
    Incorrect input processing & \textbf{169 (21.9\%)} & \textbf{71 (17.0\%)} & \textbf{79 (22.9\%)} & \textbf{28 (21.2\%)} & \textbf{152 (27.7\%)} & \textbf{297 (32.5\%)} & \textbf{249 (26.9\%)} \\
    \hline
    \textbf{Total} & 772 (100\%) & 417 (100\%) & 345 (100\%) & 132 (100\%) & 548 (100\%) & 913 (100\%) & 924 (100\%) \\ 
    \end{tabular}
\end{table*}

In contrast, \enquote{Incorrect input processing} errors are more frequent in Java (26.0\%) and Rust (22.4\%) than in C++ (4.3\%) or C (1.3\%). These differences are statistically significant across all language pairs (two-proportion z-tests with Bonferroni correction and $\alpha$=0.05). This can be explained by differences in input-handling semantics: in C and C++, the generated code frequently relies on input functions such as \texttt{cin} and \texttt{scanf}, which skip whitespace and do not explicitly recognize empty lines. As a result, certain input format mismatches silently propagate incorrect values, leading to incorrect outputs rather than runtime failures. In contrast, input handling mechanisms in Java and Rust tend to preserve line structure and enforce stricter parsing, causing unhandled input irregularities to surface as runtime errors instead.

Another notable observation is the higher prevalence of \enquote{Recursion errors} in Java (14.1\%) compared to C++ (8.3\%), C (1.2\%), and Rust (2.9\%). These differences are also statistically significant across all language pairs (two-proportion z-tests with Bonferroni correction and $\alpha$=0.05). These errors typically result from unbounded recursive calls exceeding the call stack limit. The higher frequency in Java can be explained by Java's comparatively strict default stack size limits.

Finally, \enquote{Numeric-overflow} errors appear mostly in Rust (43.9\%), with few instances in Java (5.6\%), C (4.9\%), and C++ (4.0\%). This reflects Rust's runtime overflow checks in debug mode, which detect and report overflow. In C, C++, and Java, overflow occurs silently and may later cause other errors. Listing~\ref{lst:overflow-rust} shows a representative Rust example. The exponentiation \texttt{d.pow(n)} is performed without guarding against large values, causing an overflow and triggering a runtime panic. In C, C++, and Java, the same computation would overflow silently and continue with an incorrect value. This silent propagation is a significant dependability and security risk: corrupted intermediate values can affect downstream computations, and, in security-sensitive contexts, be exploited as integer-overflow vulnerabilities. The fact that these faults go undetected in most languages suggests LLM-generated code may routinely introduce arithmetic errors that are invisible without explicit overflow checks or dedicated static analysis.

\begin{figure}[b]
\vspace{-3pt}
\begin{lstlisting}[style=smallcpp, caption={Example of a numeric overflow error.}, label={lst:overflow-rust}]
fn main() {
    ...
    let n = parts[0]; % 1000000000000
    let p = parts[1]; % 1000000000000
    ...
    if n == 1 { ...
    } else {
        for d in (1..=p).rev() {
            if p % d == 0 && (d.pow(n as u32)) <= p {
                ...
\end{lstlisting}
\vspace{-13pt}
\end{figure}

When examining problem difficulty, the distribution of runtime error types varies significantly across difficulty levels (chi-square tests with Bonferroni correction and $\alpha$=0.05).
\enquote{Out-of-bound} errors are the most frequent across all levels, growing from 35.6\% in the easiest tasks to 48.4\% in the hardest, suggesting that boundary mistakes become more likely as problems become more complex and involve more elaborate data structures or indexing logic. In contrast, \enquote{Incorrect input processing} errors are common in easier problems (25.9\%), but their relative frequency drops as difficulty rises and the relative frequency of errors of other types increases. A few other trends stand out. \enquote{Recursion errors} rise with difficulty, peaking at medium levels (11.1\% at difficulty 2), which likely reflects more frequent use of recursive algorithms in problems that require deeper search or divide-and-conquer strategies. Overall, these patterns show that, while simple mistakes dominate early on, higher-difficulty tasks tend to expose structural and logic-related issues that test the models' ability to manage complexity.

Table~\ref{tab:runtime-model-cpp} shows the distribution of Java runtime errors (chosen due to better model performance). Most errors fall into the \enquote{Out-of-bound} category, accounting for roughly 40\% across all models. Other error types show a more heterogeneous distribution. For instance, \enquote{Incorrect input processing} errors are much more frequent in smaller open-source models, occurring around 150–250 times (32.5\% in qwen2.5-coder:7b), compared to only 70–80 occurrences in larger proprietary ones (17.0\% in Gemini-2.0-Flash and 22.9\% in GPT-4.1-mini), with some of these differences being statistically significant (two-proportion z-tests over all model pairs with Bonferroni correction and $\alpha$=0.05), particularly in comparisons involving one smaller and one bigger model, while most other pairwise differences are not statistically significant. This suggests that larger models are better at understanding user requirements and translating them into correct logic.

Conversely, \enquote{Recursion errors} appear more often in larger models, with significantly higher proportions in GPT-4.1-mini (29.9\%) and Gemini-2.0-Flash (28.3\%) than in smaller models (e.g., 7.9\% in qwen2.5-coder:7b). Differences between larger and smaller models are statistically significant (two-proportion z-tests over all model pairs with Bonferroni correction and $\alpha$=0.05), while most comparisons among smaller models are not. This is caused by the lower frequency of other errors, which increases the relative share of recursion-related errors and allows execution to reach deeper recursion levels where errors occur. 

Similarly to the trends observed for compilation errors, Qwen2.5-Coder:14b produces fewer runtime errors than its general-purpose counterpart while maintaining a similar distribution of error types. This suggests that code-specific fine-tuning primarily reduces the frequency of runtime errors without substantially altering the underlying types of failures.

Table~\ref{tab:feedback-runtime} shows the impact of feedback on runtime errors. We label an error as \emph{not fixed} if the updated code still fails to compile or raises a runtime exception, \emph{partially fixed} if it runs but produces incorrect output, and \emph{fixed} if it passes all tests.

Compared to compilation errors, we observe that runtime errors are far less likely to be fully fixed. For instance, \enquote{Out-of-memory} remains unsolved in 66.5\% of cases, with only 5.1\% fully fixed, while \enquote{Recursion error} is not solved in 58.9\% of cases and only 2.6\% fixed. This contrasts with errors like \enquote{Missing import}, which appear in compilation errors and are relatively straightforward to correct once identified. The lower fix rates for runtime errors suggest that these issues often involve more complex reasoning about program behavior, such as memory usage, recursion depth, or index handling, which are inherently more difficult for models to resolve completely.

\begin{table}[b]
    \vspace{12pt}
    \centering
    \caption{Impact of feedback on runtime errors}
    \label{tab:feedback-runtime}
    \begin{tabular}{llll}
    \toprule
    Error & Not Fixed & Partially Fixed & Fixed \\
    \midrule
    Numeric-overflow & 65.2\% & 27.5\% & 7.3\% \\
    Out-of-memory & 70.6\% & 24.4\% & 5.0\% \\
    Undefined-behavior & 75.3\% & 20.1\% & 4.5\% \\
    Out-of-bound & 67.5\% & 28.1\% & 4.5\% \\
    Null pointer dereference & 65.5\% & 26.9\% & 7.6\% \\
    Recursion error & 62.5\% & 33.9\% & 3.6\% \\
    Divided by zero & 65.0\% & 30.7\% & 4.3\% \\
    Incorrect input processing & 64.8\% & 25.4\% & 9.9\% \\
    \bottomrule
    \end{tabular}
\vspace{-7pt}
\end{table}

\begin{table}[t]
    \centering
    \caption{Impact of CoT on Runtime Errors (Java)}
    \label{tab:CoT-runtime}
    \begin{tabular}{ll}
    \toprule
    Numeric-overflow & 2.6\% (-2.8\%) \\
    Out-of-memory & 3.3\% (+1.8\%) \\
    Out-of-bound & 42.8\% (-2.7\%) \\
    Null pointer dereference & 4.0\% (+0.5\%) \\
    Recursion error & 16.8\% (+2.5\%) \\
    Divided by zero & 1.5\% (-0.8\%) \\
    Incorrect input processing & 26.2\% (+0.4\%) \\
    \hline
    Total & 4090 (+39) \\
    \end{tabular}
\end{table}

Table~\ref{tab:CoT-runtime} shows the impact of \gls{CoT} prompting on runtime errors for Java. We again focus on a single language to isolate the effect of prompting from language-specific differences. Unlike in compilation errors, the shifts in the runtime error distribution are relatively small: all changes remain below 3.5\%, indicating that \gls{CoT} does not substantially alter the overall profile of runtime failures. This suggests that, runtime errors are less sensitive to prompting style and are more strongly tied to deeper flaws in program logic.

Still, some patterns emerge. The relative frequency of \enquote{Out-of-bound} errors decreases from 45.7\% to 42.3\%, while \enquote{Numeric-overflow} drops from 5.6\% to 2.6\%. A likely explanation is that \gls{CoT} prompting encourages more explicit step-by-step reasoning, which may help the model better track array boundaries and arithmetic operations, reducing simple indexing and overflow mistakes. At the same time, \enquote{Recursion error} increases from 14.1\% to 16.7\%, and \enquote{Out-of-memory} rises from 1.5\% to 3.3\%. This may reflect the fact that \gls{CoT} prompting encourages the generation of more elaborate solutions, which are more likely to rely on recursive decompositions or auxiliary data structures.

\subsection{Underlying Failure Mechanisms}

Overall, our results indicate that errors are not isolated, but instead reflect a small number of recurring failure mechanisms related to incomplete program construction, loss of structural consistency as problem complexity increases, and weak reasoning about execution and constraints.

\textbf{Incomplete program construction.}
A large fraction of compilation errors reflects the generation of incomplete programs, where required components are missing or not properly integrated. For example, \enquote{Missing import} accounts for up to 57.5\% of compilation errors in C++ and 36.1\% in C, while \enquote{Undeclared variable} and \enquote{Use of undeclared function} are also consistently observed across languages. These errors suggest limitations in the model’s ability to assemble all necessary elements of a solution and, in some cases, maintain their consistency (e.g., due to scoping issues). Overall, these errors indicate that many failures stem from missing or improperly integrated components rather than incorrect logic.

\textbf{Structural inconsistency under complexity.}
As problem difficulty increases, we observe a shift toward errors involving interactions between multiple components, with errors such as \enquote{Incompatible parameter types} (compilation), and \enquote{Out-of-bound} and \enquote{Recursion errors} (runtime) becoming more frequent. In contrast, errors that do not depend on such interactions (e.g., missing imports or incorrect input processing) often decrease. This indicates that higher complexity primarily amplifies failures in coordinating interacting components rather than uniformly increasing all error types.

\textbf{Execution reasoning and constraint handling.}
Runtime errors such as \enquote{Out-of-bound}, which accounts for up to 59.9\% of runtime errors in C++ and remains dominant across all languages, and \enquote{Incorrect input processing}, reaching 26.0\% in Java and 22.4\% in Rust, reflect failures in reasoning about program execution, particularly in handling edge cases, constraints, and input assumptions. Their prevalence across languages indicates that these are systematic limitations in reasoning about program behavior rather than isolated syntactic issues. As a result, even when code is syntactically correct, models often fail to handle boundary conditions and execution constraints, which are critical for correctness in practice.

\rn{\textbf{Contributing factors.}
Several factors plausibly explain these patterns. \enquote{Incompatible parameter types} and Rust-specific errors likely stem from Rust's stricter type system and lower training-data representation. \enquote{Incorrect input processing} is more common in Java and Rust because their input APIs raise exceptions on format mismatches, whereas \texttt{cin}/\texttt{scanf} in C/C++ silently skip whitespace, turning the same logic error into an incorrect output rather than a runtime failure. Structural errors rising with problem difficulty is consistent with longer programs, where consistency across distant parts becomes harder to maintain.}

\subsection{Overall trends}\label{sec:overall-trends}

Our analysis reveals a set of broad, practical patterns that hold across languages, models, and difficulty levels. 

\textbf{Taxonomy fit.} The \enquote{Other} category is small and language-specific labels appear where expected, supporting the taxonomy's usefulness for characterizing compilation and runtime errors across languages.

\textbf{Error distribution is heavily impacted by programming language and model.} The distribution and types of both compilation and runtime errors vary across languages and models, reflecting differences in training data, language semantics, and modeling choices.

\textbf{All models make basic mistakes.} All models frequently generate code with trivial, recurring errors (e.g., missing imports and simple type mistakes), indicating that current models are not reliable for unsupervised code generation.

\textbf{Errors are concentrated in a small set of classes.} In both compilation and runtime errors, most failures fall into few categories, meaning targeted detection and mitigation strategies, such as automated import repair or boundary validation, could yield substantial improvements with minimal effort.

\textbf{Surface-level errors are fixable; structural errors are not.} Iterative feedback resolves errors such as \enquote{Missing import} reliably, as these require only local fixes. In contrast, structural errors including \enquote{Incompatible parameter types}, \enquote{Out-of-bound}, and \enquote{Recursion errors} show significantly lower fix rates, reflecting deeper limitations in model reasoning that iterative prompting alone cannot address.

\textbf{Input handling and boundary logic are disproportionate sources of failure.} Out-of-bound errors are the most common runtime failures across all languages, and incorrect input processing accounts for a large share of failures in Java and Rust specifically. Supplying fixed input-handling scaffolding alongside problem descriptions, and systematically validating array bounds and loop conditions in code review, can eliminate entire classes of errors with minimal effort.

\textbf{High risk of undetected numeric-overflow vulnerabilities.} Rust exhibits a substantially higher incidence of numeric-overflow runtime errors than other languages due to its built-in overflow checks, which trigger panics. Arithmetic faults in unchecked languages execute silently, allowing corrupted computations to propagate undetected. This suggests that arithmetic faults in LLM-generated code may often remain hidden in languages or environments without overflow checks, posing a dependability risk and, in some contexts, a potential security concern when such code is integrated into production systems. More broadly, language choice affects error visibility, as the same generated code may appear correct in one language while containing hidden faults in another, highlighting the need for language-aware validation strategies rather than assuming correctness from the absence of a raised exception.

\textbf{CoT prompting has an impact on the  error distributions} For compilation, it reduces simple errors such as \enquote{Missing import}, but increases errors associated with more complex code, including undeclared references and attribute misuse. For runtime errors, \gls{CoT} is less impactful, with only minor shifts from simpler execution failures toward more structural issues such as recursion and memory-related errors.

\section{Threats to Validity}\label{sec:ttv}

\textbf{Internal validity.} Some error categories are underrepresented in the labeled dataset, so we cannot evaluate the classifier's accuracy for all classes. To address this, we stratified samples by programming language and randomly selected examples within each stratum, ensuring the labeled subset reflects both the language distribution and relative error frequencies in the full dataset.
In addition, error classification relies on \glspl{LLM}, so some labels may be incorrect. To mitigate this, we evaluated performance on 600 manually labeled compilation and runtime samples, confirming strong overall accuracy. \rn{\rn{A confusion-matrix-based sensitivity analysis, using per-class precision and recall from the evaluation set, confirmed that corrected distributions align closely with observed ones for all dominant classes and do not impact the conclusions.}}

\textbf{Construct validity.} The taxonomy used in this study includes both language-agnostic and language-specific categories. While this asymmetry is necessary to preserve semantic accuracy, it may affect cross-language comparability. In addition, the analyzed code-generation tasks are self-contained and do not capture the full complexity of real-world software development, such as multi-file dependencies or user interaction. To mitigate these threats, we applied a consistent labeling scheme and uniform classification criteria across all languages and models, ensuring that relative trends remain meaningful despite these limitations. Our analysis focuses on compilation and runtime errors and excludes incorrect-output cases (i.e., programs that execute but produce wrong results) and timeouts. This choice may bias the observed error distribution toward syntactic, type-related, and runtime failure modes, underrepresenting higher-level logical or algorithmic mistakes. We mitigate this risk by explicitly scoping our conclusions to errors that surface through compilation or execution failures, which remain critical for reliability and security in practice.

\textbf{External validity.} The generalizability of the findings is limited by the selection of models and languages. Although the dataset spans seven \glspl{LLM} and four programming languages, it does not include the most recent state-of-the-art models. To mitigate these limitations, this work assessed models of various sizes, architectures, and access levels, strengthening confidence that the observed trends are representative of broader \gls{LLM} behavior. Additionally, the use of competitive-programming-style problems limits generalizability, as these tasks are self-contained and do not capture aspects of real-world development such as multi-file dependencies. This setting may also amplify certain errors, particularly those related to input handling due to strict input formats. However, it provides a controlled environment that isolates fundamental code-generation behavior without confounding factors such as tooling or environment configuration. \rn{As such, the results reflect baseline capabilities likely to persist in more complex settings: missing includes affect any IDE-less project, type inconsistencies emerge in long generated code, and boundary errors arise wherever generated code indexes data structures.}

\section{Conclusions}\label{sec:conclusion}

Despite their widespread use and promise for code generation, \glspl{LLM} still exhibit significant limitations. This paper presents a large-scale empirical study of errors in \gls{LLM}-generated code, characterizing the types and prevalence of errors across models and languages. We analyze 86,726 compilation and runtime errors produced by seven \glspl{LLM} in four programming languages, examining how error behavior varies with language, problem difficulty, and iterative feedback.

Our results show that error patterns differ substantially across models and languages, reflecting differences in training data and language-specific challenges. Even bigger commercial models frequently produce simple but consequential mistakes. We repeatedly observe failures linked to weak input handling, missing boundary checks, unsafe memory assumptions, and arithmetic faults. While some languages detect these issues at runtime, others fail silently, increasing the risk of undetected security-relevant problems. Overall, \rn{although our analysis does not cover incorrect outputs,} these findings suggest current \glspl{LLM} cannot reliably produce secure, correct code without human oversight, highlighting the need for rigorous validation and review in AI-assisted development.

As future work, we plan to extend our analysis to additional languages and models, and to incorporate more complex tasks involving multi-file dependencies, external libraries, and larger codebases to better reflect real-world scenarios. We also aim to apply the same analysis to incorrect outputs and timeout errors, and investigate why certain error types persist under iterative feedback and how models modify code during refinement. We expect this work to provide a comprehensive view of the limitations of \gls{LLM}-generated code and to lay a strong foundation for targeted mitigation strategies, improved prompting techniques, and more robust code-generation models.

\bibliographystyle{IEEEtran}
\bibliography{IEEEabrv,bibliography}

\end{document}